\documentclass[a4paper,11pt]{article}
\usepackage{pos}

\title{TAMBO: A Novel Neutrino Telescope for High-Energy Astrophysical Neutrino Detection}

\author*[a]{P. Zhelnin}
\author[a]{J. Dacpano}
\author[]{C. Argüelles $^{a}$ on behalf of the TAMBO collaboration}

\affiliation[a]{Laboratory of Particle Physics and Cosmology at Harvard University,\\
Cambridge, MA 02138, USA}

\emailAdd{pzhelnin@g.harvard.edu}
\emailAdd{carguelles@g.harvard.edu}

\abstract{The detection of astrophysical neutrino point sources remains challenging due to  atmospheric backgrounds obscuring signal and statistical penalties from the look-elsewhere effect. The Tau Air-shower Mountain-Based Observatory (TAMBO) is a neutrino telescope that achieves unprecedented signal-to-background discrimination in the 1-1000 PeV energy range. Leveraging its unique deep valley geometry, TAMBO will generate an exceptionally pure neutrino sample, enabling precise investigations of neutrino sources. Preliminary sensitivity studies demonstrate TAMBO's potential to map diffuse and point-source neutrino emissions, representing a significant advancement in high-energy neutrino astronomy.}

\FullConference{
}

\begin{document}

\maketitle

\section{Introduction}
Neutrino astronomy began in 2013 with the discovery of a diffuse astrophysical neutrino flux ~\cite{IC2013}. 
Since then, the field of neutrino astronomy has advanced rapidly, yet a couple key questions remain unresolved: 
\begin{itemize}
    \item Where are all the neutrino point sources?
    \item How does the neutrino energy spectrum behave above 1~PeV?
\end{itemize} 
The rest of this proceeding will discuss the background of each of these questions and in response detail how TAMBO will aim to help answer each question. 

\section{Neutrino Point Sources}

One issue hindering neutrino point source searches is the look-elsewhere effect.
This problem arises because neutrino telescopes independently monitor many potential source locations, increasing the probability of detecting a random fluctuation that mimics signal.
To correct for this, the significance of any detected neutrino excess must be corrected downward to account for the number of independent tests via a trials factor. 
 
To reduce trials factors, the current paradigm  is to use models that predict concomitant neutrino emission with high-energy photon sources~\cite{Kheirandish_2021,Waxman_1997} -- only looking for astrophysical neutrinos from X-ray or $\gamma$-ray sources. 
While searches using high-energy photon catalogs have had some success, TAMBO can provide an alternative by offering a \textit{pure} astrophysical neutrino catalog to support point source searches at traditional neutrino telescopes.

\subsection{TAMBO Neutrino Catalog}
\begin{figure}[htbp]
  \centering
  \includegraphics[width=1.0\textwidth]{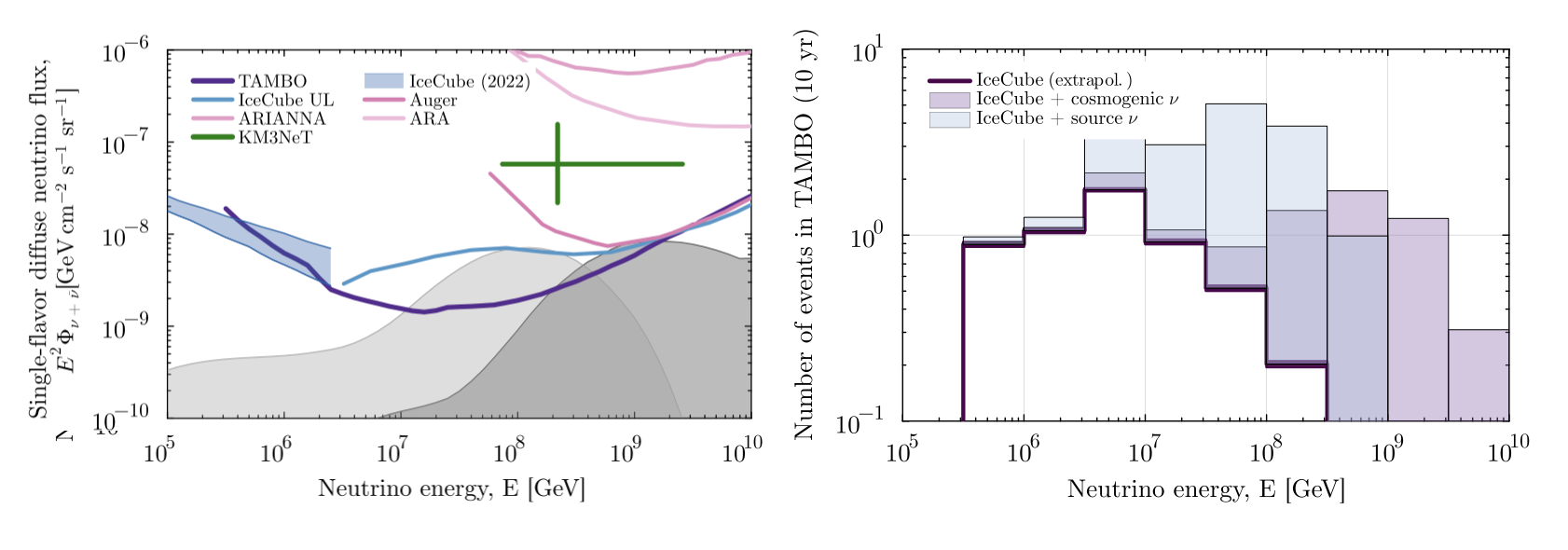}
  \caption{\textbf{Left}: Single flavor upper limits to a diffuse astrophysical flux for various experiments. The TAMBO line shows the 90$\%$ sensitivity. The darkened regions correspond to the two classes of neutrino source and cosmogenic models.  \textbf{Right}: The TAMBO full array expected neutrino events in 10 years assuming the IceCube astrophysical diffuse flux and various neutrino source and cosmogenic models. }
  \label{fig:rates}
\end{figure}

TAMBO will observe 1 to 3 astrophysical neutrino events per year, as shown in Figure \ref{fig:rates}. 
Although this rate may be lower than that of conventional neutrino observatories, the cosmic purity will be substantially higher. 
The key to this purity arises from TAMBO's unique design. 
TAMBO will deploy 5,000 detector units covering a total area of approximately $50~\mathrm{km}^2$ on one face of a \textit{deep-valley} to monitor the opposite face for decaying taus coming from Earth-skimming tau neutrino interactions. 

The valley design is crucial: the detector-less valley side provides overburden to protect the signal region from atmospheric backgrounds while also providing sufficient column density for astrophysical tau neutrinos to interact. 
The result is that any air-shower initiated by a tau decay will appear as a clear signal of an astrophysical neutrino, unlikely to be produced by atmospheric sources. 

So, for every event detected, TAMBO can report the direction with sub-degree angular resolution to neutrino telescopes, enabling targeted searches in the sky regions surrounding each reported location.
If no significant counterpart is found by these telescopes, the events reported by TAMBO can provide insights about cosmogenic neutrino production or even more exotic sources of high-energy neutrinos.

\subsection{Directional Reconstruction}
\begin{figure}
  \centering
  \includegraphics[width=0.9\textwidth]{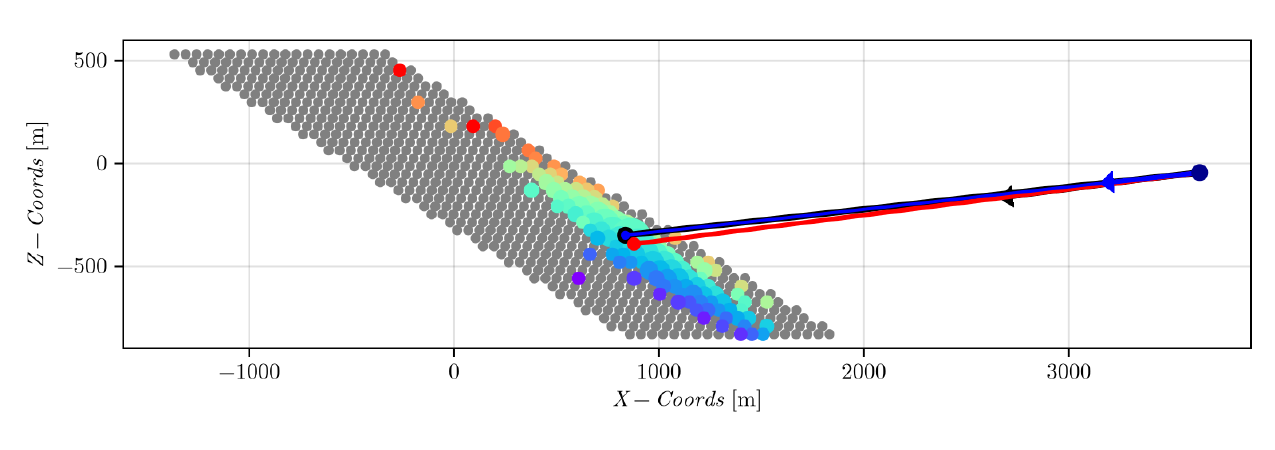}
  \caption{An example of a well-reconstructed tau-induced air-shower with a TAMBO prototype array. The black arrow shows the reconstructed direction, while the blue arrow indicates the true direction, a separation of about 0.5$^\circ$. Colored dots represent detector units that registered hits, with the dot size proportional to the number of particles detected. The color of each dot corresponds to the particle arrival time, ranging from blue (earlier) to red (later) in the shower.}
  \label{fig:reco}
\end{figure}

The effectiveness of the TAMBO neutrino catalog relies on achieving precise directional reconstruction to minimize the trials factor penalty for Cherenkov-based neutrino telescopes. 
As a result, significant effort has been devoted to characterizing TAMBO's reconstruction performance. 
Fortunately, since TAMBO is designed to detect air-showers, reconstruction techniques from cosmic-ray detectors such as IceTop can be adapted. 
Initial reconstruction results are promising, as illustrated in Figure~\ref{fig:reco}.

\section{Neutrino Spectrum Above 1 PeV}
The spectrum of astrophysical neutrinos above 1 PeV remains largely unknown. Only a handful of neutrino events have been observed above 1~PeV, and just two above 10~PeV, including the notable KM3NeT Very-High-Energy (VHE) event~\cite{VHEEvent}. Optical Cherenkov neutrino telescopes are primarily sensitive to sub-PeV neutrinos, while proposed radio-based experiments will be effective above 1~EeV, leaving a gap in sensitivity in the 1~PeV to 1~EeV energy window.  TAMBO is designed to bridge this gap: TAMBO is expected to increase the sample of supra-PeV neutrinos by an order of magnitude, thereby enabling detailed studies of the astrophysical neutrino spectrum in that energy regime and shedding light on sources responsible for supra-PeV neutrino production.

\begin{figure}[htbp]
  \centering
  \includegraphics[width=0.6\textwidth]{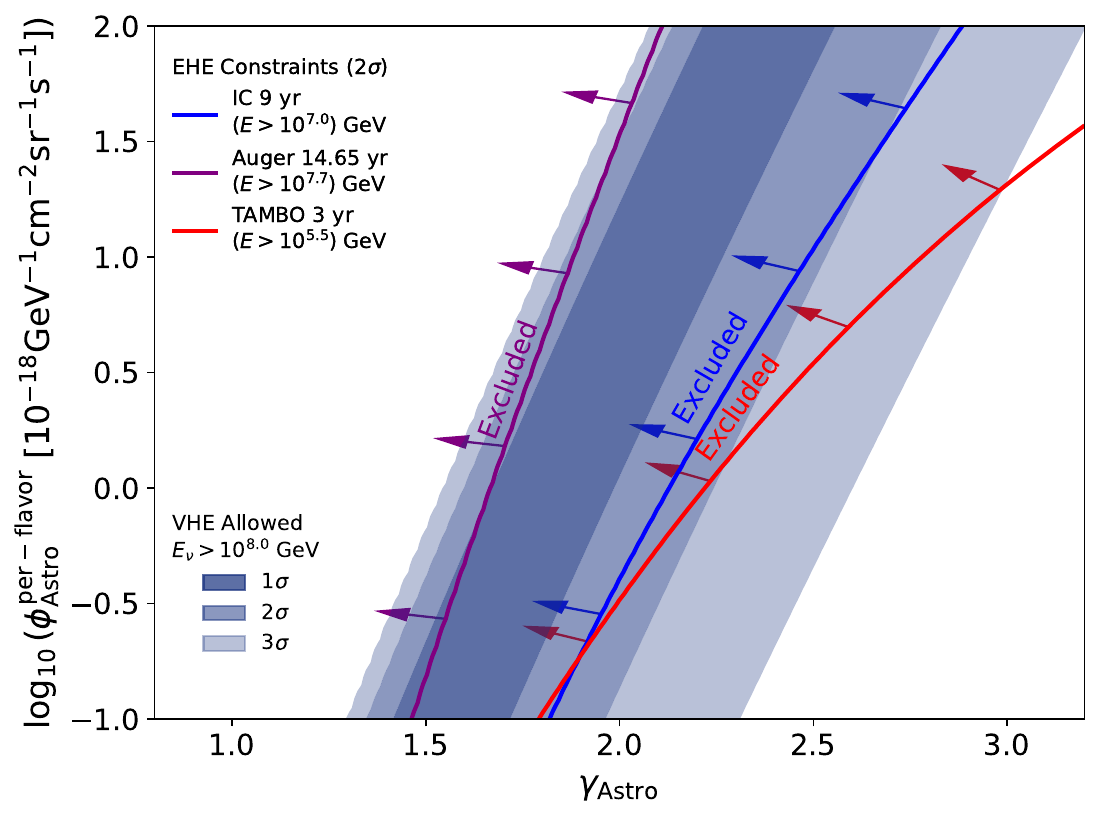}
  \caption{Fit results of KM3NeT, IceCube, Auger data and TAMBO projected sensitivity to a single power law flux model. 
  All results are fit assuming the 90$\%$ confidence level energy range of the KM3NeT VHE event. 
} 
  \label{fig:km3net}
\end{figure}

The origin of the KM3NeT VHE event, in particular, remains an open question. Owing to its unprecedented projected sensitivity to the diffuse astrophysical flux in the 1~PeV to 1~EeV range (see Figure~\ref{fig:rates}, left), TAMBO will, after three years of operation, be able to set world-leading constraints on the diffuse origins of the KM3NeT event (see Figure~\ref{fig:km3net}). This capability is especially valuable given that follow-up searches for coincident neutrino production from the sky region surrounding the KM3NeT event have so far been inconclusive \cite{VHEEvent}.

\section{Conclusion}
TAMBO’s unprecedented sensitivity in the PeV to EeV energy range will enable precise mapping of the neutrino sky and address outstanding questions regarding the origin and spectrum of high-energy astrophysical neutrinos. By bridging the observational gap between traditional optical Cherenkov and radio neutrino detectors, TAMBO is poised to advance our understanding of the most energetic processes in the universe. For a comprehensive overview of TAMBO’s design and projected capabilities, see~\cite{TAMBOpaper}.

\newpage
\section{Acknowledgments}
PZ is supported by the David $\&$ Lucile Packard Foundation. CAA are supported by the Faculty of Arts and Sciences of Harvard University, the NSF AI Institute for Artificial Intelligence and Fundamental Interactions, the Research Corporation for Science Advancement, and the David $\&$ Lucile Packard Foundation.

\bibliographystyle{unsrtnat}
\bibliography{references}


\end{document}